# Coherent Transport Through a Quadruple Point in a Few Electron Triple Dot.

L. Gaudreau[1,2], A.S.Sachrajda[1], S.Studenikin[1], P.Zawadzki[1], A.Kam[1], and J.Lapointe[1]

[1] *National Research Council, 1200 Montreal Road, Ottawa Canada K1A 0R6*
[2] *Université de Sherbooke, Quebec, Canada J1K 2R1*

**Abstract.** A few electron double electrostatic lateral quantum dot can be transformed into a few electron triple quantum dot by applying a different combination of gate voltages. Quadruple points have been achieved at which all three dots are simultaneously on resonance. At these special points in the stability diagram four occupation configurations are possible. Both charge detection and transport experiments have been performed on this device. In this short paper we present data and confirm that transport is coherent by observing a π phase shift in magneto-conductance oscillations as one passes through the quadruple point.

## INTRODUCTION

Few electron lateral quantum dots and coupled quantum dots are considered promising candidates for quantum information applications as charge and spin qubits[1]. Recently, for the first time, a few electron triple quantum dot system was realized[2] and the stability diagram mapped out using charge detection techniques[3] down to the (0,0,0) regime (where (l,m,n) refer to the number of electrons l,m,n in each of the dots respectively). In this short paper we provide evidence that transport through this system is coherent when all three dots are in resonance.

## EXPERIMENT AND RESULTS

Details of the device[4,5,6], the charge detection techniques[2] employed as well as a description of tuning procedures are provided in reference 2. In order to observe transport phenomena through the triple dot the electron arrangement was shifted by (0,1,2) from the empty (0,0,0) configuration. The three quantum dot potentials (A,B,C) were arranged in a ring with dots A and B coupled to one lead and dot C coupled to the other. A schematic of this arrangement is included in figure 1. Both DC transport experiments and charge detection measurements using an adjacent quantum point contact were performed. Figure 1 shows a zoom of the stability diagram of the triple dot in the regime of interest at settings where a quadruple point (i.e. where four electronic configurations are resonant) has been formed at point X. The regime studied in reference 2 is closer to the (0,0,0) regime and all charge transfer lines (i.e. at which electrons transfer between dots) were visible making setting the quadruple point experimentally straightforward. At the regime in figure 1, however, the charge transfer lines between dots A and B were not detectable and the quadruple points were set through a careful analysis of the line slopes, line amplitudes and the lengths of the observable charge transfer lines (bright lines in the figure). More details will be provided elsewhere. The inset shows a wider charge detection region in which the change in slope and intensity of the more horizontal lines (associated with dots A and B) at the quadruple point can clearly be observed. The left hand plot is obtained from charge detection experiments whilst the right hand greyscale is a low bias transport measurement over the same regime. The electron configurations are also shown. Note that the charge transfer processes between dots A and B (e.g. (1,1,2) to (0,2,2)) are not visible directly but are inferred from the analysis as mentioned above.

Several points need to be stressed in the data. The quadruple point is set at point X. A charge reconfiguration[2] occurs (0,2,2) to (1,1,3) as the boundary at C is crossed. Magneto-transport measurements have been performed in detail at points

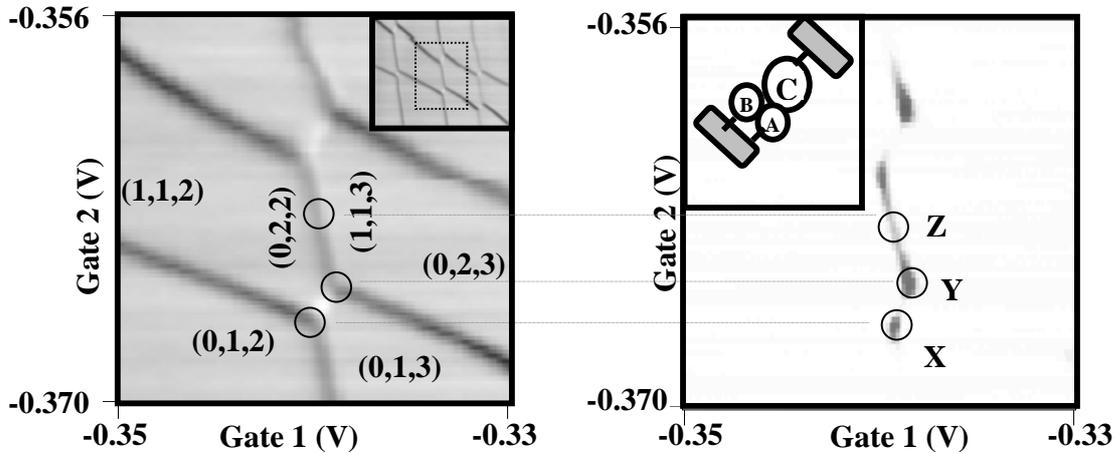

**FIGURE 1.** Charge detection and transport measurements through a quadruple point (X) in a triple dot circuit (see text for details). The insets contain a wider charge detection region and a schematic of the quantum dot circuit.

X, Y and Z. A wide variety of phenomena was observed. The phenomena include magneto-conductance fluctuations which undergo $\pi$ phase changes (as may be expected from phase rigidity[7] in two terminal measurements) and visibility modulations of these fluctuations. Figure 2 shows one example. A high bias measurement (300μeV) is made through the quadruple point passing from (0,1,2) to (0,1,3). During passage through the quadruple point (where transport occurs) two other electronic configurations also become resonant (0,2,2) and (1,1,2). A broad conductance peak with structure is observed (inset of figure 2). At either edge of the peak magnetoconductance oscillations with a period 28mT can be seen (see figure) but differing by a $\pi$ phase change. On the peak center itself a more complicated behaviour occurs involving additional periods. The fluctuations persist up to 700mK.

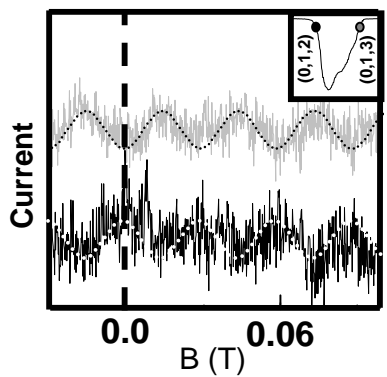

**FIGURE 2.** Magneto-conductance oscillations at a quadruple point (see text for details). Maximum peak current was 35 pA.

## CONCLUSIONS

Transport measurements have been performed on a triple quantum dot circuit, analogous to a quantum ring. Magnetoconductance fluctuations were observed at quadruple points accompanied by a variety of phenomena related to coherent transport such as $\pi$ phase changes of the Aharanov-Bohm like oscillations.

## ACKNOWLEDGMENTS

We would like to acknowledge fruitful discussions with Pawel Hawrylak, Marek Korkusinski, Aashish Clerk and Karyn Lehur. We thank Z.Wasilewski and J.Gupta for the 2DEG wafer. ASS and AK acknowledge support from the Canadian Institute for Advanced Research. ASS acknowledges funding from NSERC.